\documentclass[twocolumn]{jpsj2} %% two-column layout
\title{Precise Pressure Dependence of the Superconducting Transition Temperature of FeSe: Resistivity and $^{77}$Se--NMR Study}

\author{Satoru \textsc{Masaki}$^{1}$\thanks{E-mail address: masaki@port.kobe-u.ac.jp}, Hisashi \textsc{Kotegawa}$^{1,5}$\thanks{E-mail address: kotegawa@crystal.kobe-u.ac.jp}, Yudai \textsc{Hara}$^{1}$, Hideki \textsc{Tou}$^{1,5}$, Keizo \textsc{Murata}$^{2}$, Yoshikazu \textsc{Mizuguchi}$^{3,4,5}$, and Yoshihiko \textsc{Takano}$^{3,4,5}$}

\inst{$^{1}$Department of Physics, Kobe University, Kobe 658-8530, Japan\\
$^{2}$Division of Molecular Materials Science, Graduate School of Science, Osaka City University, Osaka 558-8585, Japan\\
$^{3}$National Institute for Materials Science, 1-2-1, Sengen, Tsukuba 305-0047, Japan\\
$^{4}$University of Tsukuba, 1-1-1 Tennodai, Tsukuba 305-0001, Japan\\
$^{5}$JST, Transformative Research-Project on Iron Pnictides (TRIP), Chiyoda, Tokyo 102-0075, Japan\\

}

\abst{
We report the precise pressure dependence of FeSe from a resistivity measurement up to 4.15 GPa. Superconducting transition temperature ($T_c$) increases sensitively under pressure, but shows a plateau between $0.5-1.5$ GPa. The maximum $T_c$, which is determined by zero resistance, is 21 K at approximately 3.5 GPa. The onset value reaches $\sim37$ K at 4.15 GPa. We also measure the nuclear spin-lattice relaxation rate $1/T_1$ under pressure using $^{77}$Se--NMR measurement. $1/T_1$ shows that bulk superconductivity is realized in the zero-resistance state. The pressure dependence of $1/T_1T$ just above $T_c$ shows a plateau as well as the pressure dependence of $T_c$, which gives clear evidence of the close relationship between $1/T_1T$ and $T_c$. Spin fluctuations are suggested to contribute to the mechanism of superconductivity.
}

\kword{FeSe, superconductivity, pressure}

\begin{document}
\maketitle

After the discovery of superconductivity at 26 K in LaFeAsO$_{1-x}$F$_x$,\cite{Kamihara} many related Fe-based superconductors have been discovered.
Among them, FeSe discovered by Hsu {\it et al.} has the simplest crystal structure formed by FeSe layers only.\cite{Hsu}
The band structure and symmetry of the superconducting gap of FeSe are similar to those of other FeAs-based superconductors.\cite{Subedi,Kotegawa}
The superconducting transition temperature ($T_c$) is 8 K at ambient pressure; interestingly, it strongly depends on pressure.
Mizuguchi {\it et al.} have measured resistivity up to 1.48 GPa, and obtained 27 K as an onset value of the superconducting transition at 1.48 GPa.\cite{Mizuguchi}
The superconducting mechanism of Fe-based superconductors is still controversial.
It is important to reveal why $T_c$ increases under pressure in order to understand the superconducting mechanism of FeSe and also of the whole Fe-based superconductors.

We report resistivity and $^{77}$Se--nuclear magnetic resonance (NMR) measurements under pressure in FeSe.
The precise pressure dependence of $T_c$ was obtained by resistivity measurements up to 4.15 GPa.
$T_c$, which is defined by zero resistance, increases nonlinearly with increasing pressure, exhibiting a plateau at approximately $0.5-1.5$ GPa.
The maximum $T_c$ is 21 K at $\sim$3.5 GPa.
The nuclear-spin lattice relaxation rate $1/T_1$ is also sensitive to pressure.
The close relationship between $T_c$ and $1/T_1$ just above $T_c$ can be observed.

A polycrystalline sample is prepared by the solid-state reaction method, as described in ref.~5.
The actual composition of the sample is FeSe$_{0.92}$ owing to a deficiency in Se.\cite{Margadonna}
Electrical resistivity measurement at high pressures was carried out using an indenter cell,\cite{indenter} and a piston-cylinder cell was used for NMR measurement at high pressures.
Electrical resistivity was measured by a four-probe method using silver paste for contact.
Note that we tried the spot-weld method for contact, but it damaged the sample.
We used Daphne 7474 for the resistivity measurement and Daphne 7373 for the NMR measurement as a pressure-transmitting medium.\cite{Murata}
Applied pressure was estimated from the $T_{c}$ of the lead manometer.
The NMR measurement was performed by a standard spin-echo method.
The polycrystalline sample was powdered for NMR measurement.

\begin{figure}[b]
\centering
\includegraphics[width=0.9\linewidth]{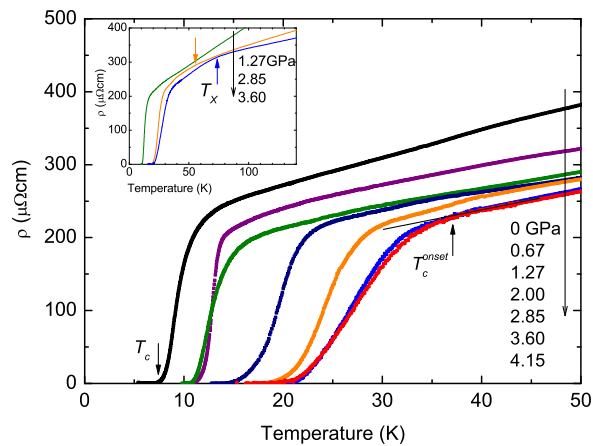}
\caption[]{(color online) Temperature dependence of $\rho$ up to 4.15 GPa. $T_c$ and $T_c^{onset}$ indicate by arrows. The inset shows $\rho$ at high temperatures. A new anomaly appears under high pressure, indicated by $T_X$.
}
\end{figure}

Figure 1 shows the temperature dependence of electrical resistivity ($\rho$) under pressure.
$T_c$ (temperature of zero resistance) increases with increasing pressure, reaching its maximum value of 21 K at 3.60 GPa.
$T_c$ slightly decreases at 4.15 GPa.
The onset temperature $T_c^{onset}$ markedly increases and reaches 37 K at 4.15 GPa.
As seen in the figure, $T_c$ is unchanged between 0.67 and 1.27 GPa.
The transition width is quite broad especially at high pressures.
The compressibility of two-dimensional FeSe is anisotropic.\cite{Margadonna2,Medvedev}
The broad transition is possibly due to the anisotropic stress on the polycrystalline sample under pressure.
The inset shows $\rho$ below 140 K.
The kink was observed at high pressures as indicated by arrows.
We denote this temperature $T_X$, but it is unknown whether or not this is the phase transition.

\begin{figure}[htb]
\centering
\includegraphics[width=0.8\linewidth]{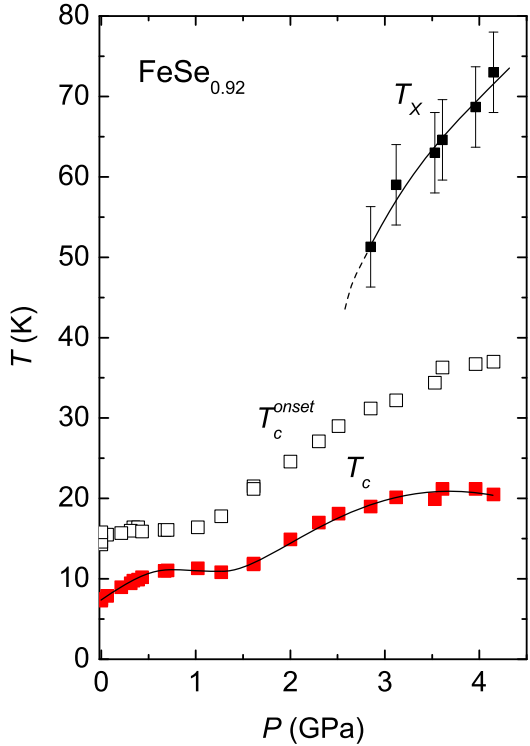}
\caption[]{(color online) Pressure-temperature phase diagram of FeSe$_{0.92}$. $T_c$ exhibits a plateau between $0.5-1.5$ GPa. The maximum $T_c$ is 21 K at approximately 3.5 GPa. The anomaly at $T_X$ appears above approximately 3 GPa.
}
\end{figure}

Figure 2 shows the pressure-temperature phase diagram of FeSe$_{0.92}$ obtained by the resistivity measurements.
$T_c$ first increases rapidly up to 0.5 GPa, and then exhibits a plateau between $0.5-1.5$ GPa.
Above 1.5 GPa, $T_c$ increases again, and becomes almost constant above 3 GPa.
On the other hand, $T_c^{onset}$ does not seem to reach its maximum even at 4.15 GPa.
The anomaly at $T_X$ appears above approximately 3 GPa, and $T_X$ increases with increasing pressure.

\begin{figure}[htb]
\centering
\includegraphics[width=0.9\linewidth]{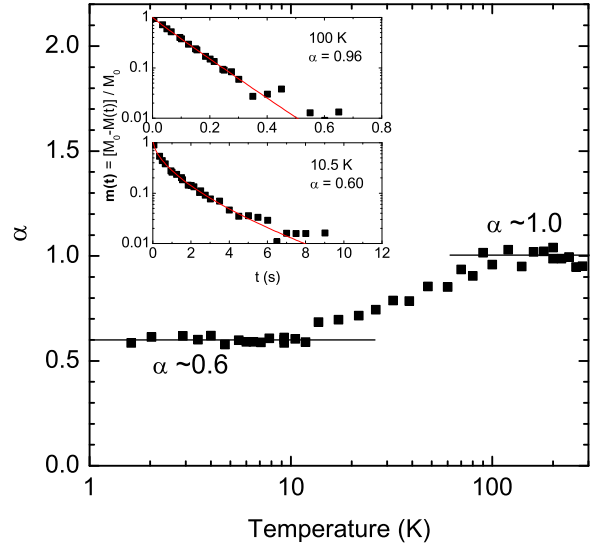}
\caption[]{(color online) Temperature dependence of $\alpha$, which is deduced by fitting the recovery curve to $m(t)=A\exp(-(t/T_1)^\alpha)$. $\alpha$ starts to decrease below $\sim$ 100 K, indicating the distribution of $T_1$. The inset shows the recovery curves at 100 and 10.5 K, fitted by different $\alpha$ values. Measurements were performed at $H=7$ T and $P=0$ GPa.
}
\end{figure}

\begin{figure}[htb]
\centering
\includegraphics[width=0.85\linewidth]{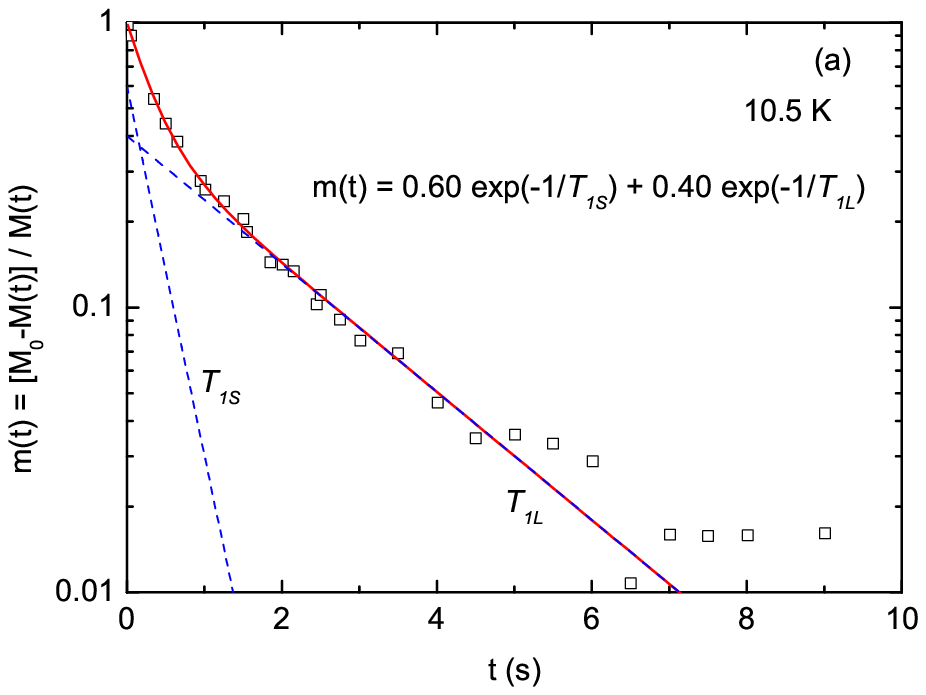}
\includegraphics[width=0.85\linewidth]{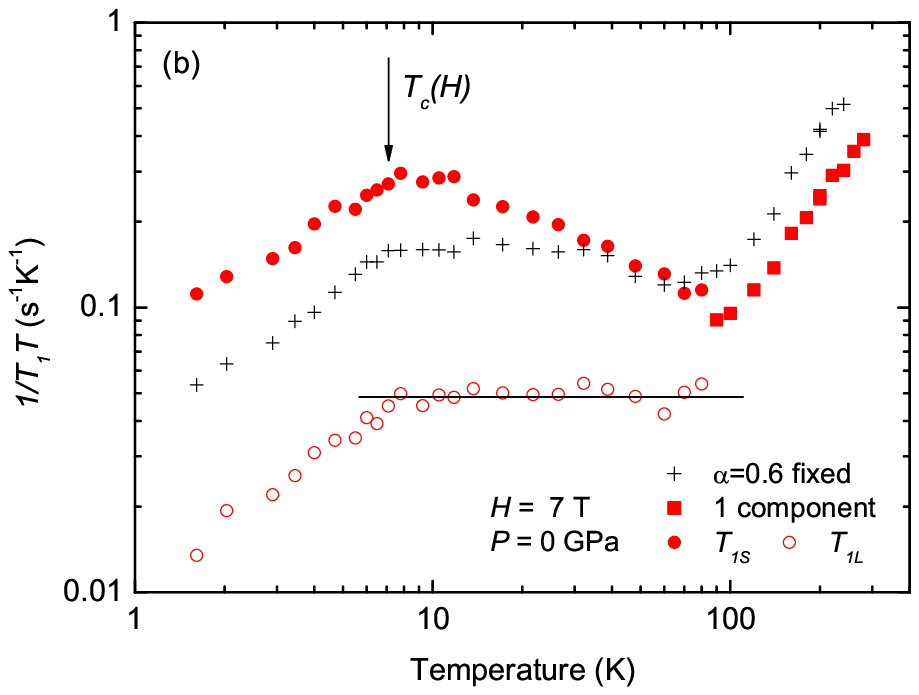}
\caption[]{(color online) (a) Recovery curve at 10.5 K. The solid curve indicates the fitting of two components as expressed in the figure. The dashed lines indicate the short and long components at $T_1$. (b) Temperature dependence of $1/T_1T$ under 7 T. $1/T_{1S}T$ increases below 100 K, indicating the development of spin fluctuations. In contrast, $1/T_{1L}T$ is temperature-independent. For comparison, $1/T_1T$ deduced by fixing $\alpha=0.6$ is also plotted (same analysis as in ref.~4).
}
\end{figure}

Next we move on the result of NMR.
We have already reported the result of NMR at ambient pressure.\cite{Kotegawa}
In our previous paper, the recovery curve $m(t)=[M_0-M(t)]/M_0$, which is needed for the determination of $T_1$, does not obey a single exponential function expected in the case of the $-1/2 \iff 1/2$ transition for the $I=1/2$ nucleus of Se.
This means the distribution of $T_1$ in the sample, but we have not drawn a conclusive remark regrading this origin in our previous paper.
Quite recently, Imai {\it et al.} have performed NMR measurements of undoped FeSe (Fe$_{1.01}$Se) at ambient pressure and under pressure.\cite{Imai}
They have reported that there is no distribution of $T_1$ in undoped FeSe.
Thus, the distribution of $T_1$ in FeSe$_{0.92}$ is considered to originate from an imperfection in the crystal structure due to Se deficiency.
In our previous paper, to determine $T_1$, we tentatively used the stretch-type function $m(t)=A\exp(-(t/T_1)^\alpha)$ while fixing the stretch coefficient $\alpha=0.6$, because the accuracy of our data at high temperatures did not allow us to treat $\alpha$ as a fitting parameter. 
However, careful measurements after the previous paper revealed that $\alpha$ actually depends on temperature.
The inset of Fig. 3 shows $m(t)=[M_0-M(t)]/M_0$ at 100 and 10.5 K.
The $m(t)$ at 10.5 K does not obey a single exponential function but obeys the stretch-type function $m(t)=A\exp(-(t/T_1)^\alpha)$ with $\alpha=0.6$.
In contrast, the $m(t)$ at 100 K obeys a single exponential function.
The temperature dependence of $\alpha$ is shown in Fig.~3.
$\alpha$ starts to decrease below $\sim$ 100 K, and shows a constant below 15 K.
Note that the analysis of $T_1$ in the superconducting state in our previous paper is not affected by the overlook for temperature-dependency of $\alpha$, because $\alpha $ shows a constant below 15 K.
However, the $T_1$ in the normal state has been misestimated because we fixed $\alpha=0.6$ up to $\sim100$ K.
$\alpha$ less than 1 indicates that $T_1$ distributes in the sample.
Although $T_1$ is expected to distribute continuously, we analyzed the $T_1$ using two-component fitting as shown in Fig.~4(a), giving the short component ($T_{1S}$) and long component ($T_{1L}$).
The volume fraction of both components is about 60\%:40\%, and it is almost temperature-independent.
Figure 4(b) shows the temperature dependences of $1/T_{1S}T$ and $1/T_{1L}T$.
$1/T_{1S}T$ increases toward a low temperature-like Curie-Weiss behavior below $\sim100$ K.
This temperature dependence is similar to that of undoped FeSe (Fe$_{1.01}$Se).\cite{Imai}
Imai {\it et al.} have suggested that this increase in $1/T_1T$ is attributed to the development of antiferromagnetic spin fluctuations from the comparison with temperature-independent Knight shift.
Thus, $T_{1S}$ is considered to originate from the region close to the stoichiometric composition.
In contrast, $1/T_{1L}T$ is almost temperature-independent in a wide temperature range, which resembles the temperature dependence of $1/T_1T$ in Fe$_{1.03}$Se.\cite{Imai}
The long component is expected to originate from the region with serious Se deficiency.
However, we stress that both components show similar power-law behaviors below $T_c(H)$, although the $T^3$ behavior seen at 2 T was not observed at 7 T.\cite{Kotegawa}
The deviation from $T^3$ behavior is considered to originate from the contribution of the vortices, but details remain unclear.
The $T_1$ distributes in FeSe$_{0.92}$, but the superconductivity seems to be homogeneous.
Above 100 K, $T_1$ is uniquely determined, and $1/T_1T$ increases with increasing temperature.
This behavior a common to some Fe-based superconductors,\cite{Nakai,Grafe} and likely originates from the effect of the band structure in the electron-doped systems.\cite{Ikeda}
The distribution of $T_1$ occurs below $\sim100$ K, which is almost consistent with the temperature of the structural phase transition from the tetragonal phase to the orthorhombic phase.\cite{Hsu,Margadonna}
The spectral weight of spin fluctuations is strongly sensitive to the Se deficiency only in the orthorhombic phase.

\begin{figure}[htb]
\centering
\includegraphics[width=0.85\linewidth]{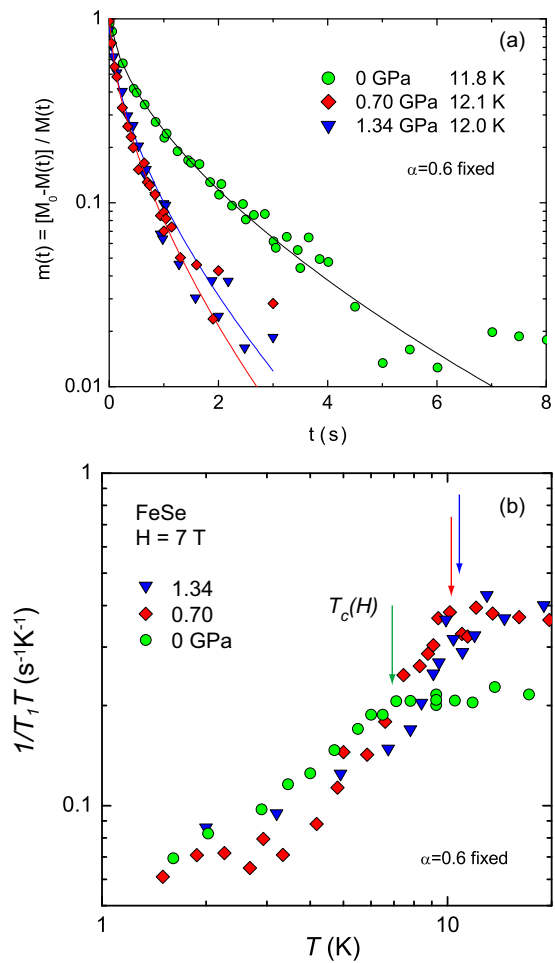}
\caption[]{(color online) (a) Recovery curves measured under pressure at similar temperatures. The solid curves show the stretch function with $\alpha=0.6$. (b) Temperature dependence of $1/T_1T$ at high pressures. Here, $T_1$ is determined by fixing $\alpha=0.6$. $T_c$ and $1/T_1T$ just above $T_c$ increase significantly from 0 and 0.7 GPa. 
}
\end{figure}

Figure 5(a) shows the recovery curves at approximately 12 K at 0, 0.70, and 1.34 GPa.
The deviation from the single exponential function is also observed under pressure.
We have tried two component fitting under pressure, but our data under pressure do not give the reliable $T_{1L}$.
Here, we discuss $T_1$ under pressure using the stretch function, because the $m(t)$ obeys the stretch function with $\alpha=0.6$ independent of pressure.
As seen in Fig.~4(b), $1/T_1T$ deduced from the stretch function gives almost average of $1/T_{1S}T$ and $1/T_{1L}T$ below $\sim15$ K, at which $\alpha$ is unchanged.
Figure 5(b) shows the temperature dependence of $1/T_1T$ under pressure.
The $1/T_1T$ is displayed below 20 K, because $\alpha$ changes significantly above $\sim15$ K.
$1/T_1T$ exhibits a drop just below $T_c(H)$ indicated by arrows.
The $T_c(H)$ increases significantly from 0 to 0.7 GPa, but the increase between 0.7 and 1.34 GPa is small.
As seen in the figure, the $1/T_1T$ just above $T_c$ ($(1/T_1T)_{Tc}$) also increases from 0 to 0.7 GPa, suggesting the enhancement of spin fluctuations.
$(1/T_1T)_{Tc}$ is almost unchanged between 0.7 and 1.34 GPa.

\begin{figure}[htb]
\centering
\includegraphics[width=0.9\linewidth]{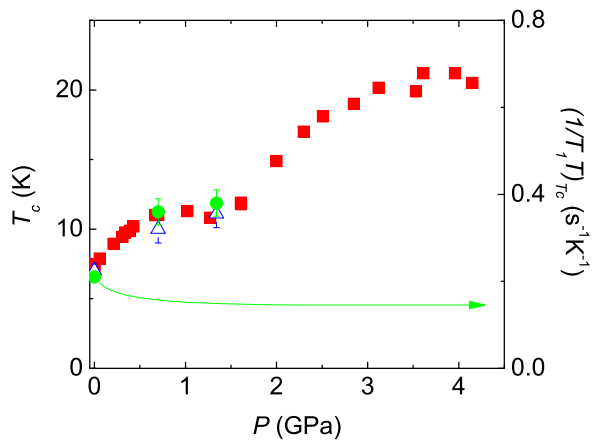}
\caption[]{(color online) Pressure-dependence of $T_c$ deduced from $\rho$ (closed square) and $1/T_1T$ (open triangle). The pressure dependence of $(1/T_1T)_{Tc}$ is also plotted (closed circle), showing a similar pressure dependence to $T_c$.
}
\end{figure}

Figure 6 shows the pressure dependence of $T_c$ deduced from $\rho$ and $1/T_1T$, and the pressure dependence of $(1/T_1T)_{Tc}$.
$T_c$'s deduced from $\rho$ and $1/T_1T$ are almost consistent with each other.
We should stress that bulk superconductivity is realized in the zero-resistance state.
As shown in the figure, $(1/T_1T)_{Tc}$ scales to $T_c$ well.
This close relationship between $T_c$ and $1/T_1$ in FeSe has already been reported by Imai {\it et al.}.\cite{Imai}
In this study, $T_c$ shows a plateau between $0.5-1.5$ GPa, and $1/T_1T$ reproduces the plateau.
This is firm evidence of the close relationship between $T_c$ and $1/T_1T$.
It is conjectured that the low-energy part of the spin fluctuations contributes to the mechanism of superconductivity in FeSe.
However, we should note that this is not a universal feature in all Fe-based superconductors.
In the LaFeAsO$_{1-x}$F$_x$ system, the antiferromagnetic spin fluctuations developing toward low temperature are not observed when the maximum $T_c$ is realized with $x=0.11$.\cite{Nakai2}

Concerning the origin of $T_X$, $1/T_1$ shows that the spin fluctuations are enhanced under pressure.
Imai {\it et al.} have reported the disappearance of paramagnetic NMR signals and the peak of $1/T_1T$ as a typical signature of a magnetic phase transition or spin freezing under high pressure.
In this context, we speculated that the anomaly at $T_X$ corresponds to a magnetic phase transition.
However, recently, Medvedev {\it et al.}, have reported that no static magnetic ordering is observed under high pressure.\cite{Medvedev}
Further investigations are needed for the elucidation.

In summary, we have investigated the pressure effect of FeSe$_{0.92}$ using resistivity and NMR measurements.
A phase diagram up to 4.15 GPa was obtained from the resistivity data.
$T_c$ increases nonlinearly with increasing pressure, reaching its maximum is 21 K at approximately 3.5 GPa.
The onset value reaches 37 K at 4.15 GPa, which is comparable to recent reports by other groups.\cite{Margadonna2,Medvedev}
However, $T_1$ shows that superconductivity is realized in the zero-resistance state as a bulk property.
The pressure effect using a single-crystalline sample is promising for a higher temperature of zero resistance.
The spectral weight of spin fluctuations distributes spatially at low temperatures in FeSe$_{0.92}$, but superconductivity seems to be homogeneous.
The pressure dependence of $1/T_1T$ just above $T_c$ reproduces that of $T_c$ well, giving firm evidence of the close relationship between $T_c$ and $1/T_1T$.
The low-energy part of spin fluctuations is suggested to contribute to the mechanism of superconductivity in FeSe.

This work has been partly supported by Grant-in-Aids for Scientific Research (Nos. 19105006, 19051014, 19340103, 19014018, 20740197, and 20102005) from the Ministry of Education, Culture, Sports, Science, and Technology (MEXT) of Japan. One of the authors (S.M.) has been financially supported as a JSPS Research Fellow.

\end{document}